\documentclass{PoS}

\usepackage[abs]{overpic}

\title{Resonance production in Pomeron-Pomeron collisions at the LHC}

\ShortTitle{Short Title for header}

\author{\speaker{Rainer Schicker}\\
        Physikalisches Institut, Im Neuenheimer Feld 226, Heidelberg University, 69120 Heidelberg, Germany\\
        E-mail: \email{schicker@physi.uni-heidelberg.de}}

\author{Roberto Fiore\\
        Department of Physics, University of Calabria, National Institute of Nuclear Physics, I-87036 Arcavacata di Rende, Cosenza, Italy\\
E-mail: \email{fiore@cs.infn.it}\\
}

\author{Laszlo Jenkovszky \\
Bogolyubov Institute of Theoretical Physics (BITP), Ukrainian National Academy of Sciences, 14-b, Metrologicheskaya str., Kiev, 03680, Ukraine  \\
E-mail: \email{jenk@bitp.kiev.ua} 
}

\abstract{A model for Pomeron-Pomeron total cross section in the resonance region 
$\sqrt{M^{2}} \le$ 5 GeV is presented. This model is based on Regge poles from
the Pomeron and two different $f$ trajectories, and includes the isolated
f$_{0}(500)$ resonance in the region $\sqrt{M^{2}}\lesssim 1$ GeV.
A slowly varying background is included. The presented Pomeron-Pomeron cross 
section  is not directly measurable, but is an essential ingredient
for calculating exclusive resonance production at the LHC.
}

\FullConference{XXIV International Workshop on Deep-Inelastic Scattering and Related Subjects\\
		11-15 April, 2016\\
		DESY Hamburg, Germany}

\begin{document}

\section{Introduction}

Central production in proton-proton collisions has been studied in the energy range
from the  ISR at CERN up to the presently highest LHC energies \cite{Albrow1}.
Ongoing data analysis include data taken by the COMPASS collaboration at the SPS \cite{COMPASS},
the CDF collaboration at the TEVATRON \cite{CDF}, the STAR collaboration at RHIC \cite{STAR}, 
and the ALICE and LHCb  collaborations at the LHC \cite{ALICE,LHCb}. 
The analysis of events recorded by the large and complex detector systems requires
the simulation of such events to study the experimental acceptance and efficiency. 
Much larger data samples are expected in the next few years both at RHIC and at the LHC 
allowing the study of differential distributions with much improved statistics. 
The purpose of the ongoing work presented here is the formulation of a Regge pole model 
for simulating such differential distributions.
 
\section{Central production}

The study of central production in hadron-hadron collisions is interesting for a variety of 
reasons. Such events are characterized by a hadronic system formed at mid-rapidity,
and by the two very forward scattered protons, or remnants thereof. The rapidity gap
between the mid-rapidity system and the forward scattered proton is a distinctive 
feature of such events. Central production events can hence be tagged by measuring 
the forward scattered protons and/or by identifying the existence of rapidity gaps. 
Central production is dominated at high energies by Pomeron-Pomeron exchange. 
The hadronization of this gluon-dominated environment is expected to produce with 
increased probability gluon-rich states, glueballs and hybrids.  Of particular interest 
are states of exotic nature, such as tetra-quark ($q\bar{q}$ + $\bar{q}q$) 
configurations, or gluonic hybrids ($q\bar{q} + gluon$).
 
\section{Central production event topologies}

The production of central events can take place with the protons remaining
in the ground state, or with diffractive excitation of one or both of the 
outgoing protons.

\begin{figure}[h]
\begin{center}
\includegraphics[width=.316\textwidth]{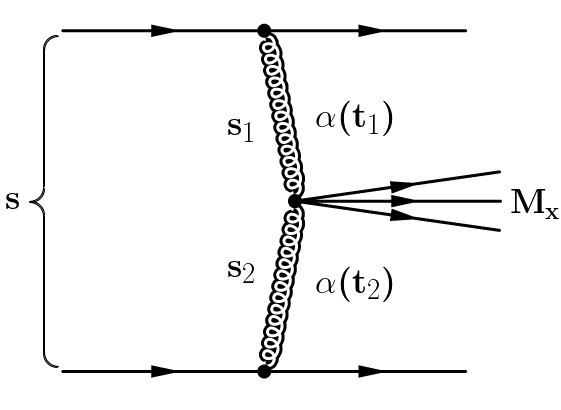}
\includegraphics[width=.286\textwidth]{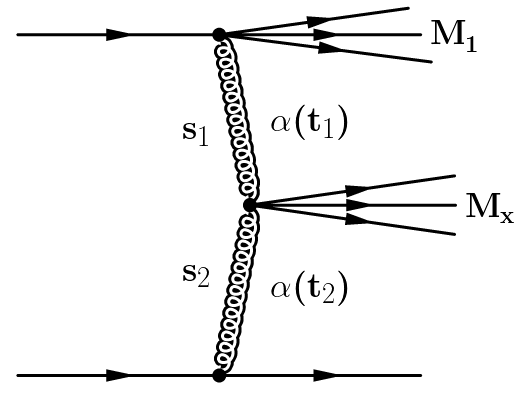}
\includegraphics[width=.280\textwidth]{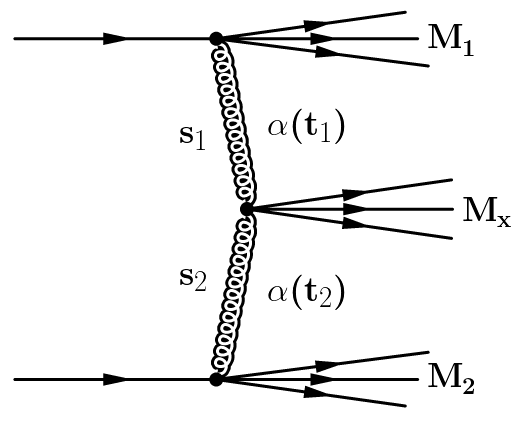}
\caption{Central production event topologies.}
\label{fig1}
\end{center}
\end{figure}

\vspace{-.2cm}
The topologies of central production are shown in Fig. \ref{fig1}.
This figure shows central production with the two protons in the
ground state on the left, and with one and both protons getting diffractively
excited in the middle and on the right, respectively. These reactions 
take place  by the exchange of Regge trajectories $\alpha(t_1)$ and $\alpha(t_2)$ 
in the central region where a system of mass M$_{x}$ is produced. The total 
energy $s$ of the reaction is shared by the subenergies $s_1$ and $s_2$ 
associated to the trajectories $\alpha(t_1)$ and $\alpha(t_2)$, respectively. 
The LHC energies of $\sqrt{s}$ = 8 and 13 TeV are large enough to provide 
Pomeron dominance. Reggeon exchanges can hence be neglected which was not 
the case at the energies of previous accelerators. 


The main interest in the study presented here is the central part of the diagrams
shown in Fig. \ref{fig1}, i.e. Pomeron-Pomeron ($PP$) scattering producing 
mesonic states of mass M$_{x}$. We isolate the Pomeron-Pomeron-meson vertex 
and calculate the $PP$ total cross
section as a function of the centrally produced system of mass M$_{x}$.
The emphasis here is the behaviour in the low mass resonance region 
where perturbative QCD approaches are not applicable. 
Instead, similar to \cite{Jenk1}, we use the pole decomposition of a dual 
amplitude with relevant direct-channel trajectories $\alpha(M^2)$ for 
fixed values of Pomeron virtualities, $t_1=t_2=const.$ 
Due to Regge factorization, the calculated Pomeron-Pomeron cross section 
is part of the measurable proton-proton cross section \cite{Jenk2}.

\section{Dual resonance model of Pomeron-Pomeron scattering}

Most of the existing studies on diffraction dissociation, single, double and 
central, are done within the framework of the triple Reggeon approach. This 
formalism is useful beyond the resonance region, but is not valid for the low 
mass region which is dominated by resonances. A formalism to account for 
production of resonances was formulated in Ref. \cite{Fiore1}. This formalism is based 
on the idea of duality with a limited number of resonances represented by 
nonlinear Regge trajectories.

\begin{figure}[htb]
\includegraphics[width=.19\textwidth]{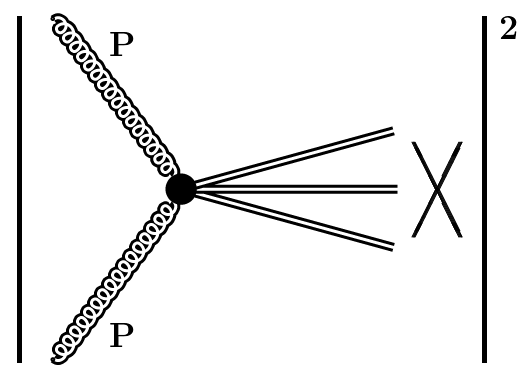}
\includegraphics[width=.038\textwidth]{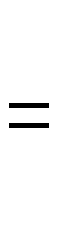}
\hspace{-0.2cm}
\includegraphics[width=.154\textwidth]{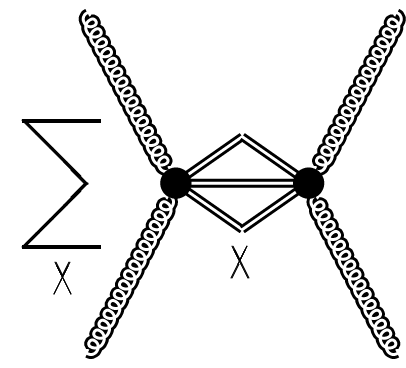}
\hspace{0.2cm}
\includegraphics[width=.037\textwidth]{pp0_0.png}
\begin{overpic}[width=.12\textwidth]{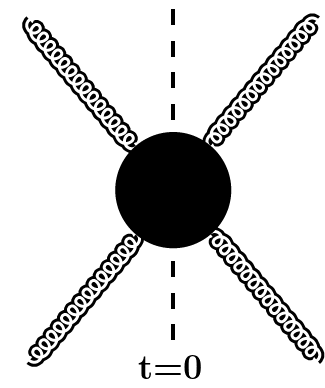}
\put(-34.,20.){\bf Unitarity}
\end{overpic}
\hspace{-0.2cm}
\includegraphics[width=.037\textwidth]{pp0_0.png}
\includegraphics[width=.132\textwidth]{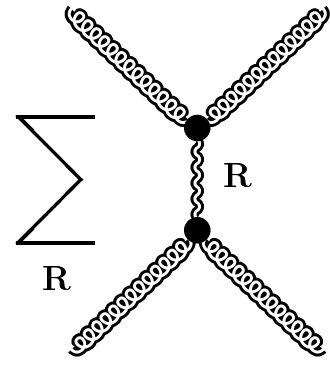}
\hspace{0.2cm}
\includegraphics[width=.038\textwidth]{pp0_0.png}
\hspace{0.2cm}
\begin{overpic}[width=.14\textwidth]{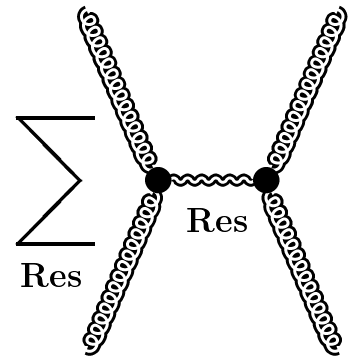}
\put(-50.,20.){\bf Veneziano }
\put(-38.,10.){\bf duality }
\end{overpic}
\caption{Connection, through unitarity (generalized optical
theorem) and Veneziano-duality, between the Pomeron-Pomeron cross section 
and the sum of direct-channel resonances.}
\label{fig2}
\end{figure}

The motivation of this approach consists of using dual amplitudes with Mandelstam
analyticity (DAMA), and is shown in Fig. \ref{fig2}. 
For $s\rightarrow\infty$ and fixed $t$ it is Regge-behaved.
Contrary to the Veneziano model, DAMA not only allows
for, but rather requires the use of nonlinear complex trajectories
which provide the resonance widths via the imaginary part of the
trajectory. A finite number of resonances is produced for limited real part of 
the trajectory. 

For our study of central production, the direct-channel pole decomposition 
of the dual amplitude $A(M_{X}^{2},t)$ is relevant. This amplitude receives 
contributions from different trajectories $\alpha_{i}(M_X^2)$, with $\alpha_{i}(M_X^2)$ a 
nonlinear, complex Regge trajectory in the Pomeron-Pomeron system,

\vspace{-0.6cm}
\begin{eqnarray}
A(M_X^2,t)=a\sum_{i=f,P}\sum_{J}\frac{[f_{i}(t)]^{J+2}}{J-\alpha_i(M_X^2)}.
\label{eq2}
\end{eqnarray}
  
The pole decomposition of the dual amplitude  $A(M_{X}^{2},t)$ is shown 
in Eq. (\ref{eq2}), with $t$ the squared momentum transfer in the 
$PP\rightarrow PP$ reaction. The index $i$ sums over the trajectories which 
contribute to the amplitude. Within each trajectory, the second sum extends 
over the bound states of spin $J$. The prefactor $a$ in Eq. (\ref{eq2}) is 
of numerical value a = 1 GeV$^{-2}$ = 0.389 mb.

The imaginary part of the amplitude $A(M_X^2,t)$ given in Eq. (\ref{eq2})
is defined by 

\vspace{-0.2cm}
\begin{equation} \label{ImA}
\Im m\, A(M_{X}^2,t)=a\sum_{i=f,P}\sum_{J}\frac{[f_{i}(t)]^{J+2} 
\Im m\,\alpha_{i}(M_{X}^2)}{(J-Re\,\alpha_{i}(M_{X}^2))^2+ 
(\Im m\,\alpha_{i}(M_{X}^2))^2}.
\end{equation}

For the $PP$ total cross section we use the norm 

\vspace{-0.6cm}
\begin{eqnarray}
\sigma_{t}^{PP} (M_{X}^2)= {\Im m\; A}(M_{X}^2, t=0).
\label{eq:ppcross}
\end{eqnarray}

\vspace{-0.2cm}
The Pomeron-Pomeron channel, $PP\rightarrow M_X^2$, couples to the Pomeron
and $f$ channels due to quantum number conservation.
For calculating the $PP$ cross section, we therefore take into account
the trajectories associated to the f$_0$(980) and to the f$_2$(1270) resonance, 
and the Pomeron trajectory.

\section{Non-linear, complex meson Regge trajectories}

Analytic models of Regge trajectories need to derive the imaginary 
part of the trajectory from the almost linearly increasing real part.
We relate the nearly linear real part of  the meson trajectory to its imaginary 
part by following Ref. \cite{Fiore2}, 

\vspace{-0.6cm}
\begin{eqnarray}
\Re e\:\alpha(s) = \alpha(0) + \frac{s}{\pi} 
PV \int_0^{\infty} ds^{'} \frac{\Im m\:\alpha(s^{'})}{s^{'}(s^{'}-s)}. 
\label{eq:disp}
\end{eqnarray}

\vspace{-0.2cm}
In Eq. \ref{eq:disp}, the dispersion relation connecting the real and imaginary 
part is shown. The imaginary part of the trajectory is related to the decay width by

\vspace{-0.6cm}
\begin{eqnarray}
\Gamma(M_{R}) = \frac{\Im m\: \alpha(M_{R}^{2})}{\alpha^{'}\:M_{R}}.
\label{eq:width}
\end{eqnarray}

\section{The Regge trajectories}

Apart from the Pomeron trajectory, the direct-channel $f$ trajectory is essential 
in the PP system.  Guided by conservation of quantum numbers, we include two 
$f$ trajectories, labeled $f_1$ and $f_2$, with mesons lying on these trajectories 
as specified in Table \ref{table1}.

\vspace{-0.2cm}
\begin{table}[h]
\begin{center}
\begin{tabular}{| c | c c | c || c | c | c ||}
\hline
& I$^{G} $& J$^{PC}$ & traj. & M (GeV) & M$^{2}$ (GeV$^{2}$) &  $\Gamma$ (GeV) \\ 
\cline{1-7}
f$_{0}$(980) & 0$^{+}$ &0$^{++}$ & $f_{1}$ &0.990$\pm$0.020 &0.980$\pm$0.040 &0.070$\pm$ 0.030\\ 
f$_{1}$(1420) & 0$^{+}$ &1$^{++}$ & $f_{1}$ &1.426$\pm$0.001 &2.035$\pm$0.003 &0.055$\pm$ 0.003\\ 
f$_{2}$(1810) & 0$^{+}$ &2$^{++}$ & $f_{1}$ &1.815$\pm$0.012 &3.294$\pm$0.044 &0.197$\pm$ 0.022\\ 
f$_{4}$(2300) & 0$^{+}$ &4$^{++}$ & $f_{1}$ &2.320$\pm$0.060 &5.382$\pm$0.278 &0.250$\pm$ 0.080\\ 
f$_{2}$(1270) & 0$^{+}$ &2$^{++}$ & $f_{2}$ &1.275$\pm$0.001 &1.6256$\pm$0.003 &0.185$\pm$ 0.003\\ 
f$_{4}$(2050) & 0$^{+}$ &4$^{++}$ & $f_{2}$ &2.018$\pm$0.011 &4.0723$\pm$0.044 &0.237$\pm$ 0.018\\ 
f$_{6}$(2510) & 0$^{+}$ &6$^{++}$ & $f_{2}$ &2.469$\pm$0.029 &6.096$\pm$0.143 &0.283$\pm$ 0.040\\ 
\hline
\end{tabular}   
\caption{Parameters of resonances belonging to the $f_1$ and $f_2$ trajectories.} 
\label{table1}
\end{center}
\end{table}

\vspace{-0.5cm}
The real and imaginary part of the $f_{1}$ and $f_{2}$ trajectories can 
be derived from the parameters of the f-resonances listed in Table \ref{table1},
and has explicitely been derived in Ref. \cite{Fiore3}.

While ordinary meson trajectories can be fitted both in the resonance and 
scattering region corresponding to positive and negative values of the 
argument, the parameters of the Pomeron trajectory can only be determined in 
the scattering region $M^2<0$. A comprehensive fit to high-energy $pp$ and 
$p\bar{p}$ of the nonlinear Pomeron trajectory  is discussed in Ref.\cite{Jenk2}

\vspace{-0.4cm}
\begin{eqnarray}
\alpha_P(M^2) = 1. + \varepsilon + \alpha^{'} M^2 - c\sqrt{s_{0}-M^2},
\label{eq:pom1}
\end{eqnarray}

\vspace{-0.2cm}
with $\varepsilon$\,=\,0.08, $\alpha^{'}$\,=\,0.25 GeV$^{-2}$, s$_0$ 
the two pion threshold s$_0$\,=\,4m$_{\pi}^{2}$, and c\,=\,$\alpha^{'}$/10\,=\,0.025.

For consistency with the mesonic trajectories, the
linear term in Eq. (\ref{eq:pom1}) is replaced by a heavy threshold mimicking
linear behaviour in the mass region of interest (M $<$ 5 GeV),

\vspace{-0.4cm}
\begin{eqnarray}
\alpha_{P}(M^2)=\alpha_0+\alpha_1(2m_{\pi}-\sqrt{4m_{\pi}^2-M^2})+\alpha_2(\sqrt
{M^2_H}-\sqrt{M^2_H-M^2}),
\label{eq:pom2}
\end{eqnarray}

with $M_H$ an effective heavy threshold $M=3.5$ GeV. The
coefficients $\alpha_{0},\alpha_{1}$ and $\alpha_{2}$ 
are chosen such that the Pomeron trajectory of Eq. (\ref{eq:pom2}) 
has a low energy behaviour as defined by Eq. (\ref{eq:pom1}).

\section{The $f_{0}(500)$ resonance}

The experimental data on central exclusive pion-pair production measured at the 
energies of the ISR, RHIC, TEVATRON and the LHC collider all show a broad 
continuum for pair masses m$_{\pi^+\pi^-} <$ 1 GeV/c$^{2}$. 
The population of this  mass region is attributed to the $f_{0}$(500).
This resonance $f_{0}$(500) is of prime importance for the understanding 
of the attractive part of the nucleon-nucleon interaction, as well as for
the mechanism of spontaneous breaking of chiral symmetry. 
In spite of the complexity of the $f_{0}$(500) resonance, and the controversy 
on its interpretion and description, we take here the practical but 
simple-minded approach of a Breit-Wigner resonance \cite{PDG} 

\vspace{-0.3cm}
\begin{eqnarray} 
A(M^{2}) = a\; \frac{-M_0\Gamma}{M^{2}-M_{0}^{2}+iM_{0}\Gamma}. 
\label{eq:BWampl}
\end{eqnarray}

The Breit-Wigner amplitude of Eq. (\ref{eq:BWampl}) is used below  for 
calculating the contribution of the $f_{0}$(500) resonance to the 
Pomeron-Pomeron cross section.   

\section{Pomeron-Pomeron total cross section}

The Pomeron-Pomeron cross section is calculated from the 
imaginary part of the amplitude by use of the optical theorem

\vspace{-0.5cm}
\begin{eqnarray}
\sigma_{t}^{PP} (M^2) \;\; = \;\; {\Im m\; A}(M^2, t=0) \;\; =  \;\; 
\sum_{i=f,P}\sum_{J}\frac{[f_{i}(0)]^{J+2}\; \Im m \;\alpha_{i}(M^{2})}
{(J-\Re e \;\alpha_{i}(M^{2}))^{2}+(\Im m \;\alpha_{i}(M^{2}))^{2}}.
\label{eq:imampl}
\end{eqnarray}

\vspace{-0.2cm}
In Eq. (\ref{eq:imampl}), the index $i$ sums over the trajectories which
contribute to the cross section, in our case the $f_{1}$, $f_{2}$
and the Pomeron trajectory discussed above.
Within each trajectory, the summation extends over the bound states
of spin $J$ as expressed by the second summation sign.
The value $f_{i}(0) =f_{i}(t)\big |_{t=0}$ is not known a priori.
The analysis of relative strengths of the states of trajectory $i$ will, 
however, allow to extract a numerical value for $f_{i}$(0) from the 
experimental data.

\begin{figure}[h]
\begin{minipage}[t]{.99\textwidth}
\begin{center}
\begin{overpic}[width=.90\textwidth]{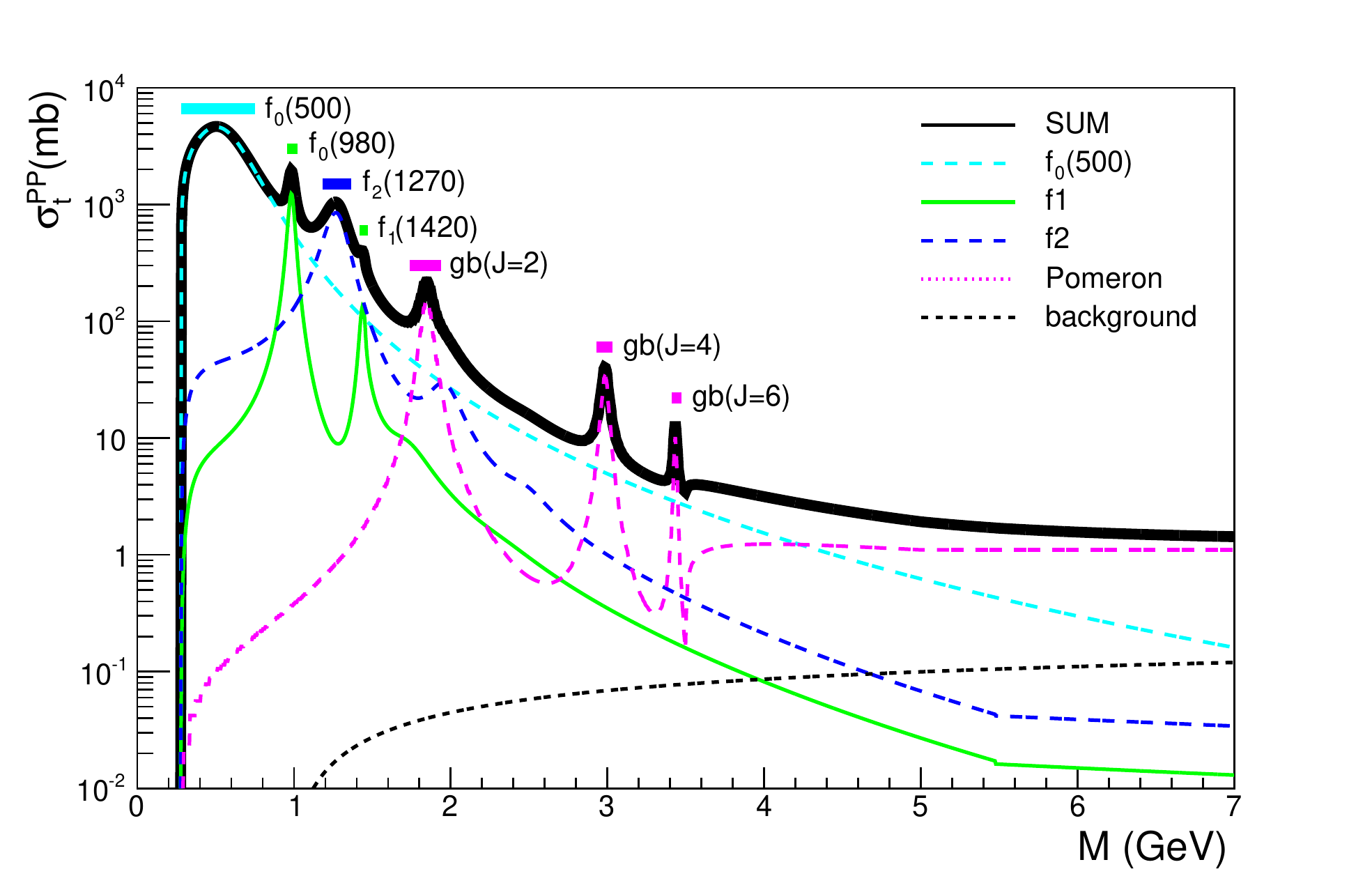}
\end{overpic}
\end{center}
\end{minipage}
\caption{Contributions of the f$_{0}$(500) resonance, the $f_{1}$, $f_{2}$ 
and the Pomeron trajectory, and of the background to PP total cross section.}
\label{fig3}
\end{figure}

The  Pomeron-Pomeron total cross section is calculated by summing over the 
contributions discussed above, and is shown in Fig. \ref{fig3}
by the solid black line. The prominent structures seen in the total cross
section are labeled by the resonances generating the peaks.
The model presented here does not specify the 
relative strength of the different contributions shown in 
Fig. \ref{fig3}. A Partial Wave Analysis of experimental data on 
central production events will be able to extract the quantum numbers 
of these resonances, and will hence allow to associate each resonance
to its trajectory. The relative strengths of the  contributing
trajectories need to be taken from the experimental data.

\end{document}